\documentclass[a4paper,11pt]{extarticle}  

\usepackage{stylesheet}

\usepackage[left=1in,top=1in,bottom=.65in,right=1in]{geometry}

\usepackage{booktabs}

\title{Incorporating climate change effects into the European power system adequacy assessment using a post-processing method}

\author{%
\textbf{%
Inès Harang,\textcolor{Accent}{\textsuperscript{1,2}} %
Fabian Heymann,\textcolor{Accent}{\textsuperscript{2,3}} %
Laurens Stoop \orcidlink{0000-0003-2756-5653},\textcolor{Accent}{\textsuperscript{4,5,6,*}} %
}\\[0.5em]
\begin{small}%
\textcolor{Accent}{\textsuperscript{1}}Mines ParisTech, Paris, France \\ 
\textcolor{Accent}{\textsuperscript{2}}ENTSO-E, Brussels, Beligium \\ 
\textcolor{Accent}{\textsuperscript{3}}University of Porto, Porto, Portugal \\ 
\textcolor{Accent}{\textsuperscript{4}}Information and Computing Science, Utrecht University, the Netherlands \\ 
\textcolor{Accent}{\textsuperscript{5}}Copernicus Institute of Sustainable Development, Utrecht University, the Netherlands\\ 
\textcolor{Accent}{\textsuperscript{6}}TenneT TSO B.V., Arnhem, the Netherlands\\ 
\textcolor{Accent}{\textsuperscript{*}}Corresponding Author: \textcolor{Accent}{laurens.stoop@tennet.eu} \\ \end{small}
}

\date{December 2020}

\begin{document}

\thispagestyle{empty}

\begin{center}
\includegraphics[width=300pt]{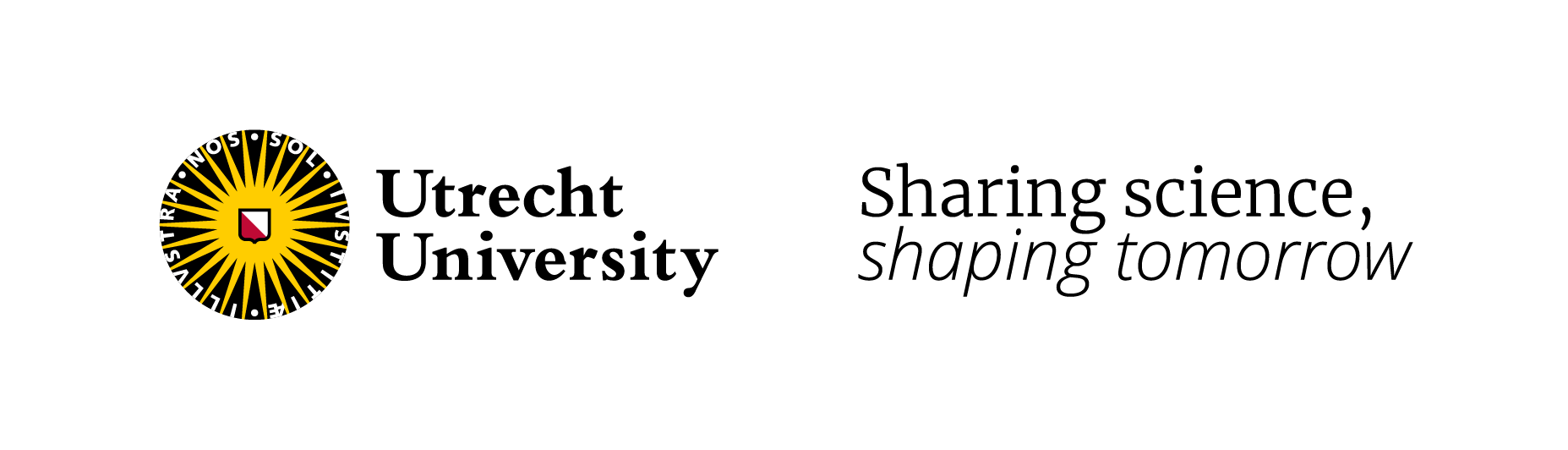}

\parbox[c]{420pt}{This is the author accepted manuscript (AAM) of the following published article:}  \vspace{20pt}

\begin{tabular}{|r|l|}\hline
\textbf{DOI} & \url{https://doi.org/10.1016/j.segan.2020.100403} \\ \hline
\textbf{arXiv DOI} & \url{https://doi.org/10.48550/arXiv.2402.17039} \\ \hline
 & \\ \hline
\textbf{author(s)} & \parbox{290pt}{Inès Harang, Fabian Heymann, Laurens P. Stoop} \\ \hline 
\textbf{title} & \parbox{290pt}{Incorporating climate change effects into the European power system adequacy assessment using a post-processing method} \\ \hline
\textbf{publication date} & December 2020 \\ \hline
\textbf{journal} & Sustainable Energy, Grids and Networks \\ \hline
\textbf{volume} & 24 \\ \hline
\end{tabular}

\vspace{30pt}
\parbox{420pt}{This AAM version corresponds to the author's final version of the article, as accepted by the journal. However, it has not been copy-edited or formatted by the journal. In addition some changes for readability have been made. \\[20pt]
This AAM is deposited under a Creative Commons Attribution--ShareAlike (CC-BY-SA) license.} 
\end{center}

\newpage

\pagenumbering{arabic}

\maketitle

\begin{abstract}
The demand-supply balance of electricity systems is fundamentally linked to climate conditions. In light of this, the present study aims to model the effect of climate change on the European electricity system, specifically on its long-term reliability. A resource adequate power system -- a system where electricity supply covers demand -- is sensitive to generation capacity, demand patterns, and the network structure and capacity. Climate change is foreseen to affect each of these components. 

In this analysis, we focused on two drivers of power system adequacy: the impact of temperature variations on electricity demand, and of water inflows changes on hydro generation. Using a post-processing approach, based on results found in the literature, the inputs of a large-scale electricity market model covering the European region were modified. The results show that climate change may decrease total LOLE (Loss of Load Expectation) hours in Europe by more than 50\%, as demand will largely decrease because of a higher temperatures during winter. We found that the climate change impact on demand tends to decrease LOLE values, while the climate change effects on hydrological conditions tend to increase LOLE values. 

The study is built on a limited amount of open-source data and can flexibly incorporate various sets of assumptions. Outcomes also show the current difficulties to reliably model the effects of climate change on power system adequacy. Overall, our presented method displays the relevance of climate change effects in electricity network studies.
\end{abstract}

\vspace{1pc}
\noindent{\it \color{Highlight} Keywords}: Power system reliability, Climate change, Resource adequacy, Electricity demand, Hydroelectric generation, Climate change impact
\vspace{1pc}

\section{Introduction}

There is now a scientific consensus that global climate will introduce substantial changes in the operation and planning of the electricity sector\autocite{bruckner2014energy}. 
This sector is tightly linked to climate change since fossil-fuel driven power generation is a cause of climate change, while transmission networks, distribution networks and the supply-side are climate-sensitive activities that are expected to be affected by climate change\autocite{bruckner2014energy,auffhammer2017climate}.

Consequently, power systems are both central in climate change mitigation (through a transition to a zero-carbon source of energy) and climate change adaptation (both on the demand side and the generation side). 
Given scientific consensus on this link between power system planning and operation and climate change and its effects, several studies have investigated the regional effects climate change would have.

\textcite{damm2017impacts} analysed the effect of climate change induced temperature variations on electricity demand using Smooth Transition Regression (STR) models. 
Their study shows that the annual energy demand in Europe would decrease, due to the reduction of the heating demand in cold days for most countries. 
On the other hand, \textcite{wenz2017north} foresees an increase of roughly 2\% in peak loads across European countries. 
This is mainly explained by the fact that from currently 5 European countries with peak demand in summer, climate change is expected to increase this number to 25 in 2100. 
The likely impact of climate change on total electricity consumption and demand peaks is also reflected in other studies outside of Europe. For example, the work of \textcite{auffhammer2017climate}, carried out in the US, predicts average electricity demand to remain steady, whereas additional peak load will require capacity additions equivalent to a 180 billion USD investment by the end of the century.

On the generation side, studies\autocite{forzieri2014ensemble,van2012vulnerability,van2013water,magagna2019water} assessed the effects of climate change on water inflow and temperature, and therefore on hydropower generation and thermal cooling. 
The main findings was a decrease of about 14\% in river flow rate in Europe (except in Scandinavia) affecting hydro generation\autocite{van2012vulnerability,van2013water}. 
Such studies suggest a decrease of hydropower potential in southern European countries would be greater than 15\%, with an average decrease of around 4\%-5\%. 
As for the thermal capacities, studies found that a rise from 1 to 2 degrees Celsius for summer water temperatures in Europe would result in a decrease of 10\% to 19\% of the usable capacity of power plants during summer periods\autocite{van2012vulnerability}.

As power generation, transmission and demand are all affected by climate change, the security of supply will be impacted as well. 
Especially the compound effects of climate change on different parts of the electricity supply chain are non-linear and therefore difficult to predict. 
Hence, its modelling is an important field of study\autocite{zscheischler2018future}. 
\textcite{turner2019compound} studied these compound impacts by assessing the effect of climate change on adequacy in the Pacific Northwest portion of the United States. 
In this study, they modelled the impact of several climate change trajectories on temperature-dependent demand patterns and hydro generation. 
Specifically to analyse the effect of compound events, events that impact various parts of an energy system simultaneously, and their affect system reliability. 
Results suggest that with climate change, tight situations during winter disappear in the analysed area. 
On the other hand, outcomes show that higher summer peak loads would coincide with lower hydropower availability, which led to a 100\% increase of summer shortfall events. 
Half of the overall climate-induced risk increases where accounted for by such compound events.

Most current power system studies, especially electricity transmission reliability studies, exclude climate change effects mostly, as the main focus is currently placed on the analysis of the effects of technology diffusion\autocite{heymann2020technology}, in particular distributed energy resources\autocite{heymann2020technology,gomes2017transmission}. 
In two recent reviews on transmission system planning\autocite{heymann2019vertical,lumbreras2016new}, climate change effects were not mentioned at all. 
One of the few studies that considered such effects was conducted by \textcite{hemmati2013state}, for a smaller region in the North-west of the US. 
In a recent report from Electric Power Research Institute\footnote{EPRI report link: \url{https://www.epri.com/research/products/000000003002014154}}, extending the scope of conventional power system analysis to include climate change is among the Top 10 challenges, without providing any additional details\autocite{turner2019compound} on how such modelling could be performed.

\subsection{The limitation of the existing studies}
Some studies in the literature focus on a regional or national scale\autocite{van2016multi}, and the few working on a European scale were not considering mitigation, or only did so with strong assumptions and relatively simple models.
\textcite{damm2017impacts} studied only the impact of the temperature factor on demand, neglecting the future penetration of technologies such as air conditioning within Europe. 
Such predictions considering different climate change pathways have been provided in\autocite{de2019households}, however this study only covers a few selected countries (in Europe: the Netherlands, Spain, France, Switzerland and Sweden). 
On the other hand, \textcite{bruckner2014energy} proposed solutions towards climate change mitigation. A selection of past research on climate change aspects linked to European power systems is shown in Table \ref{tab:ccineurope}.
However, current studies use historical climate datasets as inputs for both demand side and generation side forecast\autocite{drew2019}. 
This means that all the market modelling simulations are assuming that the future climate conditions will be similar to what we have known for the few last decades. 
In \textcite{perez2018impact}, it is assessed that the combined effects of climate-driven variation in energy demand and in water availability could result in even more important changes in power shortfall risk. 
Even though this study focused on the U.S. Pacific Northwest where the conditions are different to those of the European system, it is an indication that it is relevant to study these two effects together.

In this study, we show how to integrate climate change effects into power system adequacy studies of continental scale. 
We focus on the impact of the change in temperatures on electricity demand in Europe, and the effect of the change in water inflow (seasonality and average flows) on hydroelectric generation. 
The overall idea of this study is to provide relevant insights into how to model climate change effects in large electricity market models without mobilising full climate datasets, using a post-processing approach. 
The study analyses such effects for each of the market zones currently modelled in the European resource adequacy study, such as the Mid-term adequacy forecast (MAF), by the European Network of Transmission System Operators of Electricity (ENTSO-E). 
It contrasts different modelling variants to understand the contribution of modelling choices to security of supply levels in European electricity markets.

The main contributions of this study can be summarised as follows:
\begin{enumerate}
\item Inclusion of the effects of climate change into the European resource adequacy study (MAF). This is the first study where this has been done on a continental perimeter.
\item Development of a post-processing approach to model climate change effects based on little data input and without relying on advanced downscaling techniques.
\item A discussion on the advantages and limitations of such post-processing approaches and their use in resource adequacy studies.
\end{enumerate}

\section{Methods for modelling climate change impacts}\label{}

\subsection{Background on climate simulation models}
General Circulation Models (GCM) are commonly used to simulate the evolution and interactions of various climate parameters. 
By imposing pathways for radiative forcing, which is a result of the emission concentration pathways, climate models can project the behaviour of climate variables. 
To do so, these simulations have evolved substantially, going from static 2D representations that only consider global phenomena in the 1950s to evolving 4D representations of the earth system, see \textcite{edwards2011history}.
Modern models include various effects on a wide range of scales like ocean currents, atmospheric chemistry and ice sheet dynamics. 
The global circulation models take emission and radiative forcing trajectories as inputs, and then provide a highly granular projection of what would the climate variables look like in the future. 
The Intergovernmental Panel on Climate Change (IPCC) Assessment Reports results are based on these models. 
Note that the form of the input data has evolved to integrate more parameters\footnote{CarbonBrief on How do climate models work?  https://www.carbonbrief.org/qa-how-do-climate-models-work/}, such as the effect of emission policies and mitigation, considered in the different SRES\autocite{nakicenovic2000special} or RCP\autocite{moss2010next} trajectories.

The output climate variables from GCM's can be used as inputs for other studies, in order to assess the effect of climate change on different systems. 
In the electricity system, it is expected that climate change impacts renewable energy sources (through changing wind speeds for wind plants, water flow patterns for hydro generation, or radiation and cloud patterns for solar plants)\autocite{jerez2015impact,pryor2010climate,engeland2017space}. 
Additionally, on the thermal power plant side, it is expected that climate change may reduce the available generation capacity through the change in cooling water temperature\autocite{van2016multi} as well as rising ambient temperatures\autocite{sathaye2013estimating}. 
Furthermore, it is expected that climate change affects network operations\autocite{drew2019}, the frequencies and magnitudes of extreme weather events\autocite{vanderwiel2019extreme}, and population migration, which will then change demand patterns.

Moreover, it is expected that there will be an impact on overall electricity demand\autocite{staffell2018increasing} and hydro generation\autocite{staffell2018increasing,van2019added}. 
These last two parameters are the one which have been modelled the most. 
Therefore, this is what this study is focusing on, knowing that we will not capture the effect of all the other parameters previously mentioned.

In principle, such changes can be modelled in two ways:
\begin{enumerate}
        \item Using direct outcomes of GCM's, or,
        \item By using a form of post-analysis on existing data.
\end{enumerate}

While GCM data provides consistent projections of climate variables under different radiative forcing, it requires enhanced methodologies (e.g. downscaling of scenarios into the required spatial granularity) and costly re-analysis to project model outcomes to the regions we study, namely market zones (the smallest zones for those electricity demand, generation and exchange is modelled in MAF\autocite{entso2018mid}). 
Furthermore, the breakdown from average annual values to a more detailed time domain (e.g. hourly resolutions) requires further assumptions that, for the nature of climate change forecasts of 100 years ahead, may induce significant errors. 
Since using direct outputs of climate simulations would be costly in time and complexity, we focused our methodology on a post-analysis strategy.

\begin{landscape}
        \vspace*{\fill}
            \begin{table}[ht]
                \centering
                \caption{Power-System related climate change effects in Europe.}
                \label{tab:ccineurope}
        \begin{tabular}{p{0.45\textheight} @{\hspace{1em}} p{0.45\textheight} @{\hspace{1em}} p{0.45\textheight}}
                \hline
                Reported impact of climate change on demand & \multicolumn{2}{l}{Reported impact of climate change on hydrogeneration}  \\ \hline
                  & Impact on inflows/water variables & Impact on hydrogeneration \\ \cline{2-3}
                - Both total electricity consumption and peak loads are affected\autocite{wenz2017north} &- Average decrease inflows of 13\%-15\% by 2040\autocite{forzieri2014ensemble,van2012vulnerability}, (apart from Scandinavia) & - Average usable power plant capacity decreases by 10\%-19\% during summer due to higher mean water temperatures\autocite{van2012vulnerability,van2013water} \\
                - Temperature changes show a strong spatial heterogeneity especially between northern to southern European countries\autocite{damm2017impacts} & - Mean summer water temperatures increase by 0.8-2.3 ◦C by 2040\autocite{van2012vulnerability} & - Hydropower potential in southern countries decreases by more than15\%, with an average decrease of around4\%-5\%\autocite{van2013water}\\
                - In most European countries, CC decreases mean higher electricity demand\autocite{damm2017impacts} & - River flows decrease in South-Eastern Europe, especially in summer; whereas they may increase in other regions, especially in winter\autocite{magagna2019water} & - water scarcity of the power system in Poland, Czechia, Bulgaria, Germany, France, and Romania, where strong links to water-energy exist\autocite{magagna2019water} \\
                - European countries peaking in summer will increase from 5 to 25 by2100\autocite{wenz2017north} & - The frequency and intensity of drought swill rise, especially in Southern Europe\autocite{forzieri2014ensemble,magagna2019water} & \\
                - Peak loads may increase by roughly 2\% by 2050\autocite{wenz2017north} & \\
                  \bottomrule
        \end{tabular}
        \end{table}
        \vspace*{\fill}
\end{landscape}

\subsection{The impact of temperature changes on electric load}\label{Ext1:load_reprofiling}
In the aforementioned European adequacy study methodology, demand time series are generated using a modelling tool used for mid-term adequacy studies at ENTSO-E, see\autocite{operator2019electricity} for a more detailed outline of the framework.

\begin{figure}[ht]
        \centering
        \includegraphics[width=\textwidth]{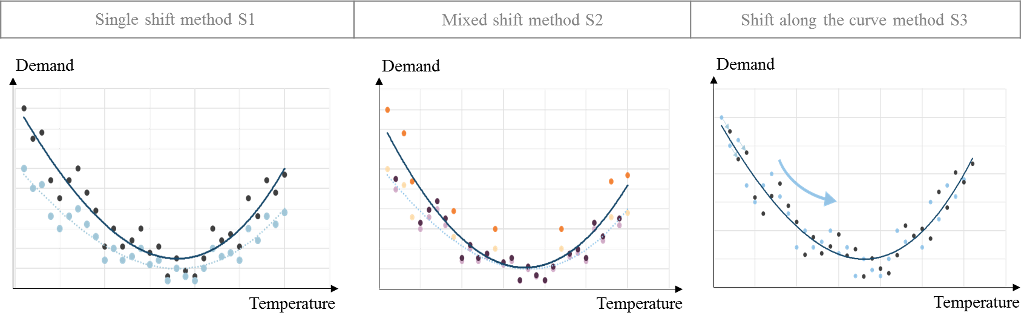}
        \caption{Temperature-load reprofiling and scaling methods to incorporate climate change. Dark blue colours show the original data and load-temperature curve, data points and curves changed by a method in other colours. The scale of this figure is arbitrary.}
        \label{fig:load_reprofiling}
\end{figure}

MAF studies do not currently consider regional or highly granular values of temperature but use aggregated and population weighted data grouped per market zone. 
In addition, in such power system studies, load patterns and their evolution are modelled on an aggregated level, neglecting the spatial heterogeneity of load growth\autocite{heymann2018emerging} and other impacting factors such as technology diffusion, for example distributed energy resources\autocite{heymann2020adopter}, or policy changes\autocite{heymann2019orchestrating}. 
Therefore, in this work, we identified three different ways to assess the effect of climate change on demand using a post-processing approach. 
In doing so, we focus on what can be achieved with available data (namely without access to any hourly demand time series of temperature under future climate scenarios for each market node). 
A graphical representation of the methods discussed below is shown in Figure \ref{fig:load_reprofiling}.

\subsubsection{Shifting the load with a single aggregated value (S1)}
The first solution (S1)
(see Figure \ref{fig:load_reprofiling})
would be to start from the results found by \textcite{wenz2017north} and to shift the load along one axis. 
As a first approximation, we could apply the same aggregated coefficient on each value of the hourly demand time series we would use “without climate change” effect. 
This is the simplest post treatment operation. 
One of the main problems of this method is that the winter days and summer days are treated the same way, where the warming of winter days should lower demand according to the sensitivity curve, which should not be the case if we start from a curve in which the adaptation of the system is already calculated\autocite{wenz2017north}. 
As a consequence, this would change the shape of the temperature sensitivity curve. 
Another issue is that this method does not consider extreme events.

\subsubsection{Shifting the load with multiple aggregated values (S2)}
A second solution (S2), Figure \ref{fig:load_reprofiling}, similar to the previous one but with an added feature, would be to consider the hours corresponding to the daily peak loads and the other hours of each day differently. 
This would require some data analysis of the time series in order to identify the hours on which to apply time-of-day specific coefficients but would offer a better management of the peak loads and therefore of extreme events. 
However, this method does not solve the problem of the winter days and the change in the shape of the sensitivity curve.

\subsubsection{Deduct load based on a temperature shift (S3)}
To preserve the temperature sensitivity curve, and, moreover, use it to consider winter and summer days differently, a third solution S3, Figure \ref{fig:load_reprofiling}, would be not to change the load independently by applying a coefficient but to shift the load along the curve starting from a temperature change. 
Here, the data treatment is more complex since we need to compute the coefficients of the regression of the temperature sensitivity curve, for each market zone, and then move each load along the curve. 
The disadvantage of this method, besides the complexity, is that we do not consider extreme events, because we use a single aggregated value of temperature change.

All three methods are summarised in Table \ref{tab:reprofiling}.
For this study, we decided to use the solution that is the most consistent with data used in the reference study case, which is the third solution (the reference sensitivity curves used in this study will be explored in more details in Section \ref{sec:methodExt1}). 
By shifting the temperature and deducting a new load, we keep the same thermal sensitivity and therefore remain consistent with the study carried out without consideration of climate change effects. 
However, the reader should bear in mind that we do not precisely consider extreme temperature events, since we are using an average value of temperature change for each market zone.

\begin{table}[hb]
        \centering
        \caption{Summary of the advantages and drawbacks of the load reprofiling methods.}
        \label{tab:reprofiling}
        \begin{tabular}{lccc}
                \hline
                Aspect & S1 & S2 & S3 \\ \hline
                Easy to implement & $+$ & $-$ & $- -$ \\
                Inclusion of extreme events & $- -$ & $+$ & $- -$ \\
                Management seasonality & $- - $ & $- -$ & $+$ \\
                Preservation of temperature sensitivity & $- -$ & $- -$ & $+$ \\ \hline
        \end{tabular}
\end{table}

\subsection{Impact of climate change on hydroelectric generation}\label{ext1:HydroLit}

Hydroelectric generation is affected by climate change through water inflows going into hydro power plants\autocite{van2013water}. 
In European adequacy studies (such as MAF), different hydropower generation technologies (pumped hydro, reservoir hydro, runoff river) are considered. These are modelled either using electricity market price signals and balancing requirements or historical river/reservoir water inflows. 
In the current model layout, ENTSO-E's adequacy studies do not consider climate change effects on the availability of water resources nor the interaction between electricity and water networks.

In our study, we model the new inflows under a scenario with climate change by modifying two parameters that have been identified as significant in the literature: the average variation of total water flows in catchments (for each country)\autocite{van2013water}, and the shift of seasonality in their profiles\autocite{forzieri2014ensemble}. 
We used an approximation under which over a full year, a relative change in total inflow is proportionally correlated to a similar change in total hydro generation. 
Time series that represent the hydrological availability have been affected following a standardised process:
\begin{enumerate}[label=\Alph*)]
        \item Determine the hydrological resource profile;
        \item Extracting simplified profile;
        \item Rescaling;
        \item Reprofiling.
\end{enumerate}

Note that the reservoirs operations will be indirectly affected as well since an optimisation of the reservoir fill and draft operations is performed in a market modelling tool used at ENTSO-E.

\section{Methodology and input data}\label{sec:methodExt1}

The overall idea of this study is to perform a post processing modification on the inputs of the European adequacy assessment study. 
The following sections will provide details on the model used in this European study (Section~\ref{Ext1:metMAF}), and then more specifically on the input files that will be modified to incorporate climate change effects (Section~\ref{sec:load_reprofiling} \& \ref{sec:hydromethod}). 
The last step consists in feeding the European adequacy study with these new inputs (Section~\ref{Ext1:metDATA}), and how we can analyse if adequacy is affected by climate change on a European scale (Section~\ref{Ext1:metAdequacy}).

\subsection{The European adequacy model}\label{Ext1:metMAF}
An adequacy study is an evaluation of the security of supply in scenarios in which resources and demand are evolving. 
The goal of such studies is to support stakeholders in their decisions of investments and help identify risks, by focusing up to a few years ahead forecast. 
Here, we are using the base case reported in 2019 for the target year 2025 (in this study labelled “MAF 2025”\autocite{entso2018mid})

\begin{figure}[ht]
        \centering
        \includegraphics[width=\textwidth]{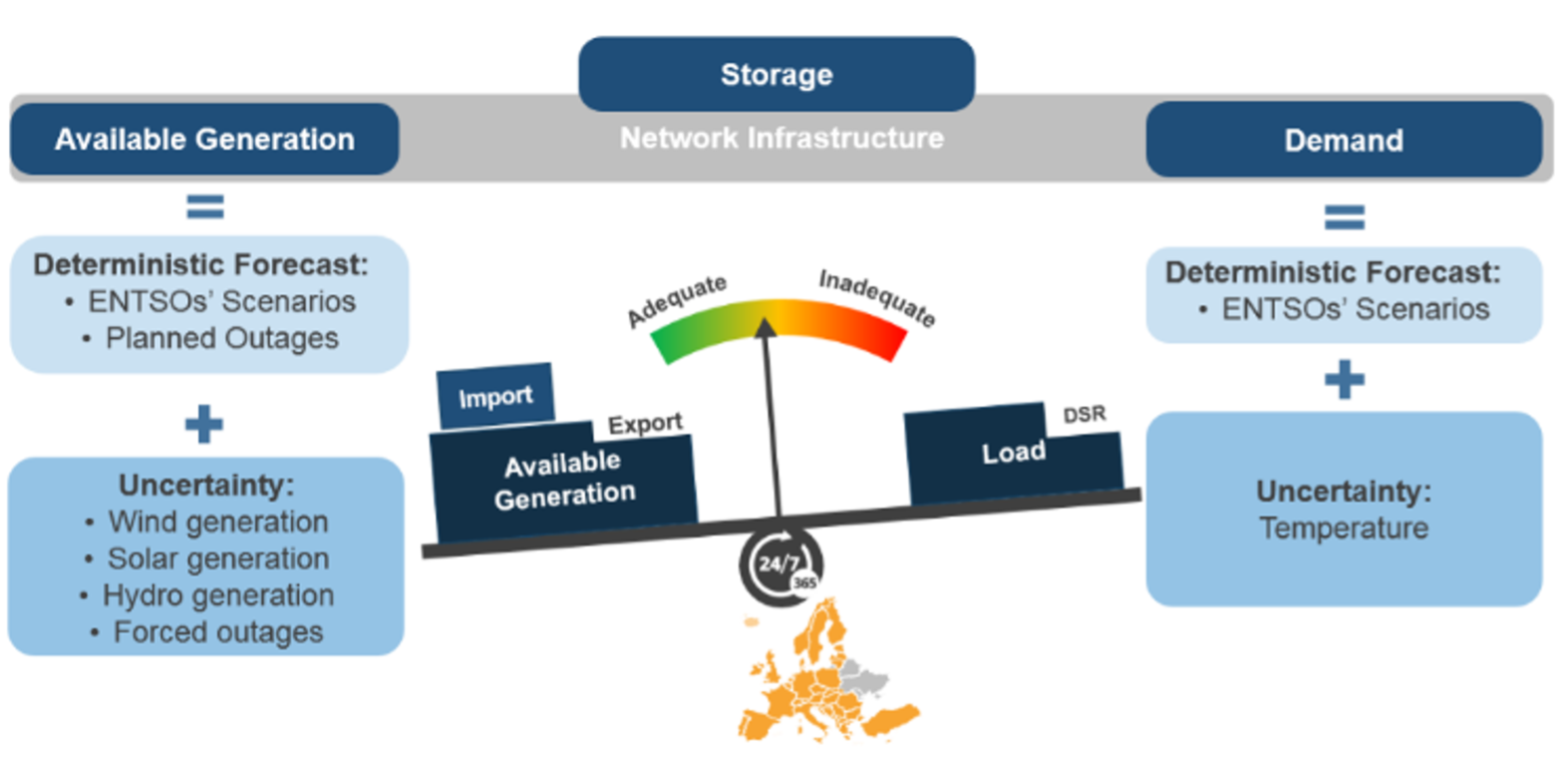}
        \caption{Illustration of the evaluation of the security of supply done in the MAF studies\autocite{entso2018mid}.}
        \label{fig:maf_studies}
\end{figure}

MAF studies combine deterministic scenario forecasts (generation, demand, planned generator outages) with random patterns (forced generator outages) and a historical, multi-temporal data-set that describes climate conditions and electricity demand\autocite{entso2018mid}. 
To do so, a random sample of certain years is created, in which the random patterns are combined with deterministic, historical climate patterns (past historical climate conditions including temperature, irradiation, wind speed, river water flows). 
As such, MAF uses Monte Carlo (MC) sampling with different combinations of these variables as shown in Figure \ref{fig:maf_studies}.
Therefore, the indices resulting from the MAF studies assess the adequacy of the generation-transmission system (Hierarchical Level 2) as defined by \textcite{li2013reliability}.
In this study, this model has been implemented in an electricity market modelling tool (here PLEXOS) that is used in MAF's process as explained in \textcite{entso2018mid}. 
Such a large-scale study on climate change effects on adequacy throughout the whole European continent, using publicly available data, has never been performed before.

\subsection{Methodology to modify demand files}\label{sec:load_reprofiling}

The demand forecasts are a result from an ENTSO-E proprietary tool\autocite{li2013reliability}. 
Forecasts built on a regression tool that accounts for the load-temperature sensitivity using historical time series. 
Load forecasts are additionally adjusted with scenarios on the uptake of electric vehicles, heat pumps, batteries and energy efficiency measures.

To simulate the effect of temperature variation on demand, we used a regression on the temperature sensitivity load curve (load against temperature). 
As the baseline we used the time series that are publicly available\autocite{entso2018mid}\footnote{ENTSO-E (TF TRAPUNTA) Demand forecasting methodology}, we took into account the evolution of the electricity sector structure without having to make strong assumptions. 
Then, we shifted the load along the regression curve of each market node by imposing a temperature variation, as outlined in Figure \ref{fig:load_reprofiling} and explained in Section~\ref{Ext1:load_reprofiling}.

\begin{figure}[h]
        \centering
        \includegraphics[width=\textwidth]{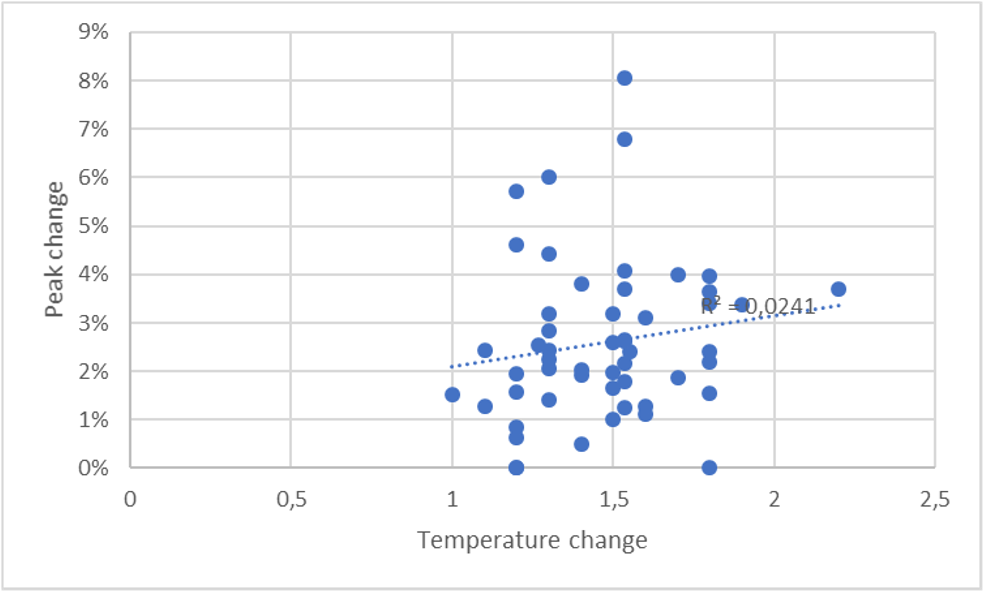}
        \caption{Correlation between change in peak load (in \% in absolute values) and change in temperature (in C) for each market node.}
        \label{fig:correlation_load-temp}
\end{figure}

The result of this shift is a new time series in the same format (hourly granularity and 35 climate years) but with a consideration of climate change as a post-treatment. 
We can observe that for most market nodes (39 out of 53), the new overall peak load over the 35 climate years becomes lower, in line with previous work\autocite{clarke2018effects}. 
This can be explained as for most countries, the peaks are in winter, and we have made the winters warmer so less demanding in heating. 
The only countries where this is not the case are southern European countries (Cyprus, Greece, Hungary, Malta, Turkey and most market zones in Italy), and the 3 market nodes for which we have not shifted the load (LUF1, LUB1 and NON1) as these showed temperature-independent load sensitivity. 
This is in the line with the work of \textcite{damm2017impacts}.
If we look at total energy demand over the year, we have similar trends: 36 market nodes out of 53 have a lower average consumption. 
For Spain, Portugal and the northern Italian market node, the peak is usually in winter, but most of the consumption is still in summer, so in average, a rise in temperature will result in a higher overall electricity demand.

If we look at the correlation between change in peak loads and change in temperature in Europe, see Figure\ref{fig:correlation_load-temp},
it can be observed that there is no clear linear relationship between the two, which can be explained by the fact that we used a non-linear shift along a 2\textsuperscript{nd} order sensitivity curve. 
However, there is a tendency in which the more the temperature changes, the more the peak load will change.

\subsection{Methodology to modify river inflows}\label{sec:hydromethod}
\begin{figure}[ht]
        \centering
        \includegraphics[width=\textwidth]{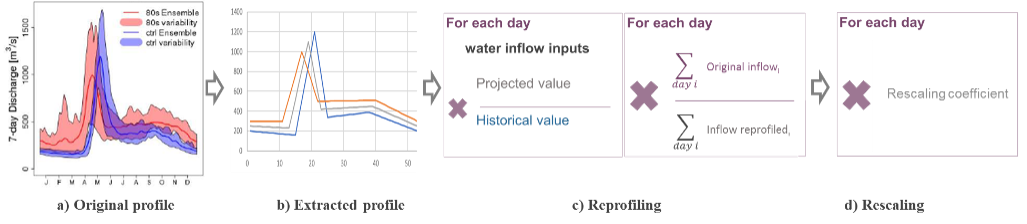}
        \caption{Schematic of the method for reprofiling and rescaling of hydrological resource availability}
        \label{fig:hydromethod}
\end{figure}

For hydrological resource availability, the idea is to work in two steps: ``reprofiling'' and ``rescaling''. 
Reprofiling is changing the seasonality of the input inflow following a trend found as found in literature, but without affecting the total inflow. 
Rescaling is changing the total inflow in the year accordingly to the data found in literature as explained in Section~\ref{ext1:HydroLit}. 
A summary of the method is shown in Figure \ref{fig:hydromethod}.

\subsubsection{Reprofiling}
For reprofiling the following three steps have been used.

First, profiles are extracted from literature.
Since we had no access to the output datasets of \textcite{forzieri2014ensemble}, on which we based our methodology, we needed to reconstruct the data by hand from the charts available in \textcite{forzieri2014ensemble}. 
To do so, we wrote a script doing the following: for a given set of coordinates of peaks (its value and the corresponding time), the scripts builds the profiles connecting the peaks one to another, in a weekly time series format. The idea is, from pictures of graphs, reconstruct the data in order to exploit it.

Second, the reprofiling coefficients are calculated. 
With the extracted data we have access to coefficients representing the new weight of a week in terms of inflow, from the control period to the projected years. 
For each week, the coefficient would be the ratio of the value of inflow in the projected years -- as approximated in the 1\textsuperscript{st} step -- to the value of inflow in the control period.

Third, rescaling is applied in order not to change the total inflow -- which has only to be modified by the rescaling phase. 
To do so, we simply multiply all the values of the time series by a constant coefficient which is a function of the average after this reprofiling operation and the average before, that we want to restore.

\subsubsection{Rescaling}
For rescaling two possibilities have been initially considered. 
Option 1 shifts all the inflow curve up or down accordingly to the scaling coefficient we use as an input. 
This would not deform the curve but only shift the average.
Option 2 applies the scaling coefficient on each value of the time series of inflow.

In both cases the average would be changed according to the percentage we use as an input for this rescaling. 
The advantage of the 1\textsuperscript{st} method is that you do not deform the profiles. 
The drawback is that it can potentially lead to negative river inflow values. 
To avoid this problem, we decided to use the 2\textsuperscript{nd} method. Since this method will change the shape of the profiles, we rescale after having reprofiled, to make sure that the reprofiling would be performed on the historical profiles.

\subsection{Input data for the MAF-model}\label{Ext1:metDATA}
To assess the effect of climate change on adequacy, we needed to modify some input files of the market modelling tool: adapting the demand input file according to temperature modifications and changing the hydro input file according to the impact of global warming on water inflows. 
The adequacy study we are considering here is the MAF for target year 2025\autocite{entso2018mid}.

\subsubsection{Data used for demand}
The demand time series are obtained by the software used within ENTSO-E\footnote{Namely the package TRAPUNTA, see for details the Demand forecasting methodology (2019) \url{https://eepublicdownloads.entsoe.eu/clean-documents/sdc-documents/MAF/2020/Demand_forecasting_methodology_V1_1.pdf}.}. 
This uses singular value decomposition to train and forecast load profiles models depending on various climate conditions. 
In addition, it identifies temperature-dependent and temperature-independent load shares and uses them to forecast adjustments, using scenarios of market evolution (penetration of heat pumps, of electric vehicles, \ldots). 
This information about the evolution of the electricity market affecting demand is collected by ENTSO-E in what is called the Pan European Market Modelling Data Base (PEMMDB). 
The final demand file is an hourly forecast, for each market node and a given target year, of electricity load. 
To be more precise, the tool produces 35 different demand time series, corresponding to 35 years of historical climate data, on which there will be a Monte-Carlo simulation performed in the market modelling tool (see \ref{sec:load_reprofiling}).
For demand, it is the same as saying we forecast what would be the electric load with the same climate as in 1982-2016 but with different electricity system and needs. 
Note that population weighted temperature values are used to represent market zones, and not just average values.

The information used to apply a ``climate change'' modification to the demand time series is an intermediate and aggregated temperature result from \textcite{damm2017impacts}, in order to have the average variation of temperature for each European county in a given target period (here 2016-2045). 
These aggregated values were calculated from temperature data of the EURO-CORDEX climate simulations\autocite{jacob2014euro}, based on the RCP8.5.

Therefore, our results should not be considered as exact estimates of power system reliability behaviour but rather capture the upper boundary of climate change effects that may be observed in the limited time horizon to 2025. 
For the countries in which there are several markets zones modelled in the ENTSO-E modelling framework (e.g. Italy, Luxembourg), the same value of temperature variation was used. When the temperature was not calculated by \textcite{damm2017impacts}
(i.e. for Albania, Bosnia and Herzegovina, Switzerland, Cyprus, Greece, Montenegro, Macedonia, Malta, Serbia, Turkey and Ukraine), we used an average value of the temperature variation of the neighbouring countries.

\subsubsection{Data used for hydroelectric generation}

In our work, we manipulate the hydrological river run-off time series stored in PEMMDB. 
These files contain, for each market zone, historical data of 35 climate years of cumulated natural inflow into reservoirs and run-off river hydro generation. 
This data is collected centrally by ENTSO-E from all its European member TSO’s as part of the PEMMDB. 
The hydrological data contains mainly weekly observations, except for the run-of-river where the temporal granularity of the information collected is daily data.

Then, we used external data for the post-treatment of these hydro files to consider effects of climate change. 
Namely, for the hydrological resource availability, we took information from two studies: \textcite{forzieri2014ensemble} and \textcite{van2013water}.

\textcite{forzieri2014ensemble} was used for reprofiling the change in inflow profiles.
This study provides profiles of discharge in 11 different stations in Europe for a ‘control period’ (1961-1990) and for projections (2071-2100). 
To modify MAF 2025, we used an intermediate value. Those projected profiles were calculated using data from climate simulations, based on the Intergovernmental Panel on Climate Change Special Report on Emissions Scenarios (IPCC SRES) A1b\autocite{nakicenovic2000special}. 
To reprofile the inflows of all the market nodes (most of which did not have a station and thus no data was available in the study), we used the data of the nearest station.

\textcite{van2013water} was used as the data source for rescaling the average change of total water inflow per country.
We considered the percentage of change in hydro potential similar as the associated change of water inflow. 
Those percentages were calculated for the period 2031-2060, based on the IPCC SRES B1 scenario. 
Since we are modifying inputs for MAF2025, this rescaling represents the higher end of climate change effects. 
A similar methodology as for temperature was used in the case where a country had several market zones, we took the same value for all of them, or for those which had not been considered by \textcite{van2013water}, we used the average value of the neighbouring countries.

\subsection{Key metrics to assess adequacy}\label{Ext1:metAdequacy}
To access adequacy in the electricity system and compare this study with a reference case of MAF 2025 (mid-term adequacy forecast for the target year 2025), we are analysing a few metrics measuring how often the electricity supply does not meet demand and how much energy it represents. 

The first metric for that is the Loss of Load Expectation (LOLE), which represents the number of hours in the target year in which generation and imported energy cannot meet demand. 
The LOLE is the average value of the unserved energy hours among all the sample years used in the Monte Carlo (MC) market model by, which are all the combinations of 35 climate years and 20 samples (S) of random draws for unplanned outages. 
LOLE is defined mathematically by:
\begin{align}
        LOLE &= \frac{1}{N} \sum_{j \in S} LLD_j
\end{align}
where $LLD_j$ is the loss of load duration ($j \in S$) of the MC simulation, and N is the number of MC simulations.

Another metric we will evaluate is the Expected Energy Not Served (EENS). 
It represents the total amount of energy in a given year that could not be supplied during the hours mentioned above. 
The EENS is the average value of unserved energy among all the sample years. 
EENS is defined mathematically by
\begin{align}
        EENS &= \frac{1}{N} \sum_{j \in S} ENS_j
\end{align}
where $ENS_j$ is the loss of load duration ($j \in S$) of the MC simulation, and N is the number of MC simulations.

\section{Results}
\begin{figure}[ht]
        \centering
        \includegraphics[width=\textwidth]{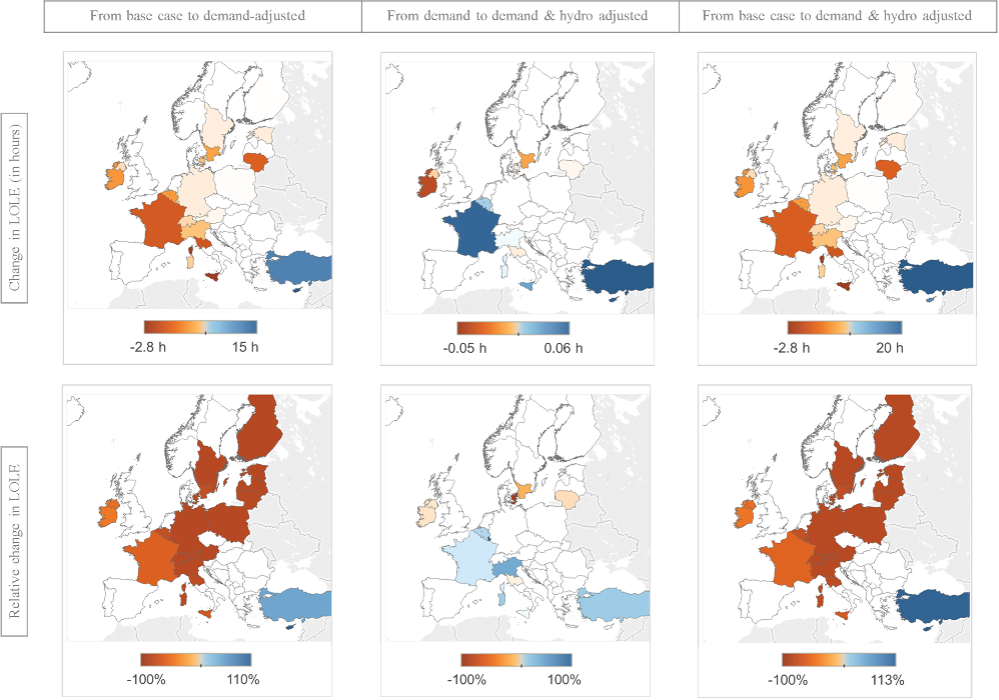}
        \caption{Change in LOLE in models MAF2025 (the base case); MAF2025 (demand CC affected) and MAF2025 (demand and hydro CC affected).}
        \label{fig:change_LOLE}
\end{figure}

In the following sections, different scenarios will be compared (see Figure \ref{fig:change_LOLE}): the reference scenario with no effects of climate change (scenario-ref) will be compared to a scenario in which the effects of climate change on demand have been considered (scenario-demand), and to a scenario in which the effects of climate change are considered on both demand and hydro (scenario-dem-hydro).
Section~\ref{Ext1:resDem} compares scenario-ref with scenario-demand,
Section~\ref{Ext1:resHydro} compares scenario-ref with scenario-dem-hydro, and finally
Section~\ref{Ext1:resComb} compares scenario-demand with scenario-dem-hydro. 
The interest is to assess the contribution of both effects, and the compound effect, on adequacy.

The interpretation of the comparison requires knowledge on the relative change in the data fields used, for this we refer to Figure \ref{fig:change_param}.

\subsection{Consideration of climate change effects on demand}\label{Ext1:resDem}

As a first step and for the sake of comparison, we will first analyse the change in adequacy in the MAF model 2025, thus only considering the effect of climate change on demand (and not hydro generation yet). 
In this run, we changed the demand input data in the market modelling tool using the modified demand time series, while keeping all other inputs constant.

The main result of this simulation is that compared to a scenario without climate change, the average LOLE among the different market nodes in Europe would decrease by 59\%; the unserved energy by 30\%.
This general trend is related to the fact that under climate change, energy demand generally decreases in Europe. Besides, there are some market nodes for which the change would be greatly significant, see Figure \ref{fig:change_param}.
For instance, the market nodes for which the increase in LOLE is maximal (Turkey, Cyprus) would see a change of 60\% and 112\% of their LOLE respectively, and 86\% and 138\% of the unserved energy. For some market nodes, the LOLE would decrease by 100\%: there would be no more hours without energy served. 
This would be the case for Austria, Switzerland, Germany, Estonia, Finland, Latvia, Poland and one market node of Sweden (SE03).

\subsection{Climate change effect on demand and hydro generation}\label{Ext1:resHydro}

Now, we analyse the change in adequacy in the MAF model 2025 if we consider the effect of climate change on demand and on hydro generation. 
In this run, we changed the demand and hydro input data in the utilised market modelling tool to the modified versions, while keeping all other inputs constant.

The main result of this simulation is that compared to a scenario without climate change, the average LOLE among the different market nodes in Europe would decrease by 56\%; the unserved energy by 11\%. 
Besides, there are some market nodes for which the climate change effects on electricity demand and hydro power generation, and thus, system reliability, would be significant,
see Figure \ref{fig:change_param}.
For instance, the market nodes for which the increase in LOLE is maximal (Turkey, Cyprus) would see a change of 110\% and 113\% of their LOLE respectively (and 150\% and 138\% of the unserved energy). 
Under this scenario of combined modelling of climate change effects on hydro generation and electricity demand, in some market nodes, the LOLE would decrease by 100\%: there would not be any more hours with energy not served,
see Figure \ref{fig:change_param}.
This would be the case for Austria, Switzerland, a market node of Denmark (DEK1), Germany, Estonia, Finland, Latvia and one market node of Sweden (SE03).

\subsection{The contribution of hydro change in the previous study}\label{Ext1:resComb}
\begin{figure}[hb]
        \centering
        \includegraphics[width=\textwidth]{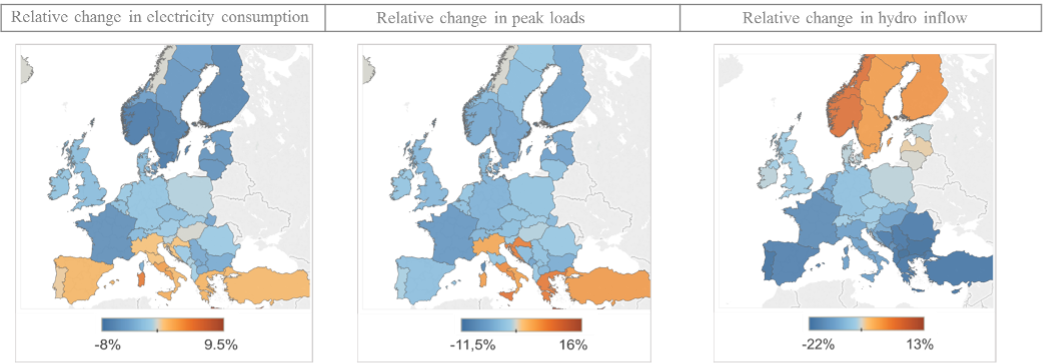}
        \caption{Change in the parameters used as inputs to the models that consider climate change.}
        \label{fig:change_param}
\end{figure}

To analyse the contribution of the change in hydro, we compare the results of the studies in which the effects of demand have been considered and in which the effects of demand and hydro are incorporated. 
From a scenario only considering the effects of climate change on demand to a scenario considering the effects on hydro as well, the average LOLE among the different market nodes in Europe would increase by 7\%; the unserved energy by 27\%; and the electricity price would decrease by 2.4\%. For specific market zones, see Figure \ref{fig:change_param}.
This shows that the contribution of hydro to the overall modelling of climate change tends to put the electric system under stress, and therefore to increase the LOLE and EENS. 
Note that we did not observe any case for which the LOLE would switch from 0 to a positive value.

\section{Discussion}
In this contribution we explored possible ways to incorporate climate change effects in power system modelling. 
Furthermore, we applied a post-processing approach to estimate the effects of climate change on the reliability levels in the European power system.

As major take-away, this work showcases the two main difficulties of modelling climate change effects in power system studies:
\begin{enumerate}
        \item The creation of forward-looking, consistent data-sets on future climate conditions, their downscaling to finer spatial and temporal resolutions as well as the treatment of model uncertainties remain major challenges.
        \item In contrast to climate data derived from measurable historical observations, strong discrepancies in future-looking climate data-sets due to different GCMs and CO2 emission scenarios make deterministic validation of forecasts intractable.
\end{enumerate}

This may require moving towards modelling using probabilistic density functions with accurate representation of extreme events rather than creating a large set of equally weighted Monte-Carlo scenarios.

It is without doubt that the chosen approach represents a simplification of all possible effects climate change could have on power system adequacy, including compound events or various likely climate change pathways. 
As the climate scenarios have similar effects of climate change in short run, this should not invalidate the results of the work presented. 
Additionally, the scenarios built in this study are consistent with the reference MAF 2019 study and build on a common dataset that includes all European TSOs. 
Specifically, by construction, the temperature sensitivity curves of demand have been preserved, and all the parameters of the MAF model are collected in a standardised process through the PEMMDB.

However, it should be mentioned that the coincidence of low hydro generation with peak temperature driven demand is not considered and may not reflect the adequacy situation in each European member state. 
In addition, the results here are indicative for an average year. Historic extreme events like the extreme cold spell from early 1985 are likely reasonably presented in this study but and any shift in frequency or magnitude of extreme events is not considered. 
To accurately assess extremes under climate change, an approach similar to \textcite{vanderwiel2020}
should be used.

Future work should, when possible be based on a consistent set of assumptions to better capture possible linkages in climate change effects, i.e. following the method described by \textcite{zscheischler2018future}, in which the importance of a consistent dataset is highlighted by the correlation of the effects of climate change on intertwined elements of the electricity system. 
As mitigation strategies involve large shares of renewable energy resources, these should be modelled and considered as well. 
Additionally, by calculating both hydro generation and electricity demand directly on GCM output variables, no assumptions need to be made on the effect of climate change on these variables. 
Furthermore, when multiple scenario’s and GCM’s are used the uncertainty of the specific scenario and climate model used can be quantified and better understood, resulting in better projections.

In our work, we forecast power system reliability under climate change. 
As the future is yet uncertain, it was not possible to measure the accuracy of our predications.

As one major contribution of this work is the inclusion of the climate change effects into a Pan-European electricity market model, our methodology has focused on comparing our simulation results to the base model (not considering CC effects). 
In fact, from a power system planner’s perspective, it is the uncertainty range beyond his baseline that is the fundamental input to the decision-making process of electricity network planning.

\section{Conclusion}
A new post-processing method was developed to model the effect of climate change on power system adequacy, through the analysis of the Loss of Load Expectation metric. 
The presented method captures climate change induced changes in electricity demand and hydrological generation patterns, where the latter is modelled using river discharge/flows.

Using this new method on the data used for the European Adequacy Assessment 2019, with the horizon towards year 2025 (MAF2025 - scenario year), the study compares future reliability situation in Europe under climate change and without modelling climate change effects.

Due to the climate change effect on temperature, a lower demand is found for most countries in Europe. 
The Loss of Load Expectation for Europe would decrease by 59\% if climate change induced demand shift is taken into account. 
A large range, from -100\% to 112\% of relative change in LOLE is observed for the individual market zones. 
When in addition to the climate change effects on demand climate change affected hydro generation is taken into account, a similar conclusion can be drawn.
An overall reduction of 56\% of LOLE is found, but there is a large range for individual market zones (100\% to 113\%).

When the two methods of including climate change are compared, we find that including both energy demand and hydro generation changes are significantly affected. 
Climate impacts on energy demand reduce the stress of the system, while climate induced impact on hydro generation induce stress in the system.

In the developed methodology, temperature variations are considered constant over the year: both the changes in winters and summers are modelled equally. 
As such, we are neglecting shifts in extreme temperature events which lay outside the scope of this analysis. 
The future integration of the drift of extreme weather events would improve the reliability of retrieved results.

Finally, the presented analysis is the first attempt to include the modelling of climate change effects in a large continental scale industrial power system study, namely the MAF study from ENTSO-E. 
The presented post-processing approach can be used as a transitory methodology to model climate change effects, until the full merging of climate and energy system studies.


\section*{CRediT Author Statement}
\textbf{Inès Harang:} Formal analysis, Investigation, Methodology, Visualization, Writing - original draft, Revision. \textbf{Fabian Heymann:} Conceptualization, Methodology, Supervision, Writing - original draft, Review \& revision. \textbf{Laurens P. Stoop:} Methodology, Conceptualization, Writing - original draft, Review, Revision \& editing.

\section*{Acknowledgments}
John Fazio, Massoud Jourabchi and Daniel Hua are kindly acknowledged for their reviewing work and constant support. 
Vincent Deschamps is also thanked for his review and advice. 
Ladislas Quazza is kindly thanked for his reviewing work on the final version of the paper. 
The authors would like to thank Nils Müller for his help in the study of water inflows, use of market models, and the System Development section at ENTSO-E for valuable feedback. 
L.P. Stoop received funding from the Netherlands Organisation for Scientific Research (NWO) under grant number 647.003.005.
The content of this paper and the views expressed in it are solely the author’s responsibility, and do not reflect the views of TenneT TSO B.V., ENTSO-E or its members.


\printbibliography

\appendix
\beginsupplement
\newpage
\section{Additional figures }\label{app:ext1}

\begin{figure}[h]
        \centering
        \includegraphics[width=.8\textwidth]{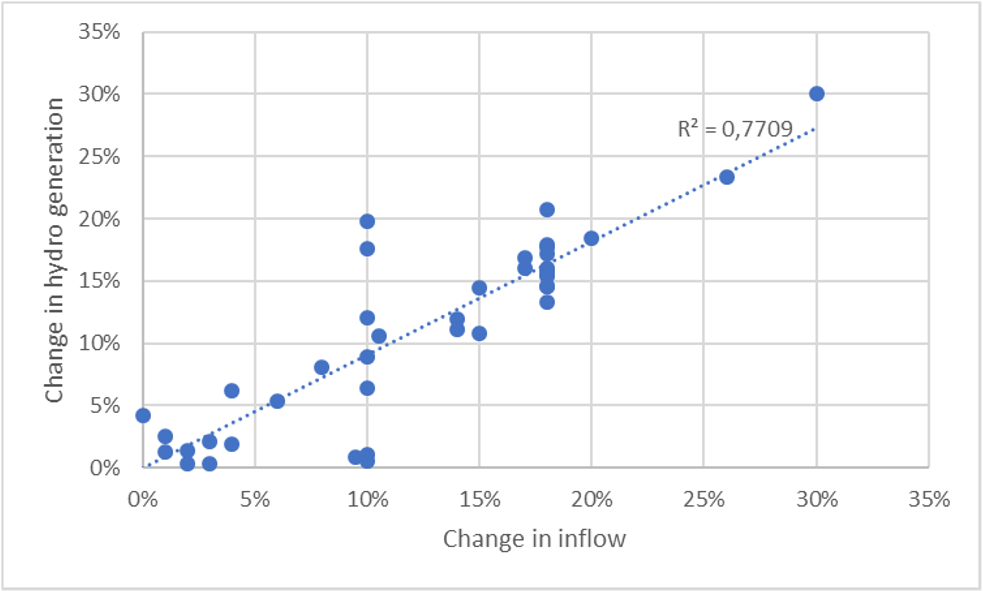}
        \caption{Correlation between hydro generation (in \% in absolute values) and percentile change in inflow for each market node. }
        \label{SIfig:corr_hydro-inflow}
\end{figure}

\begin{figure}[h]
        \centering
        \includegraphics[width=.8\textwidth]{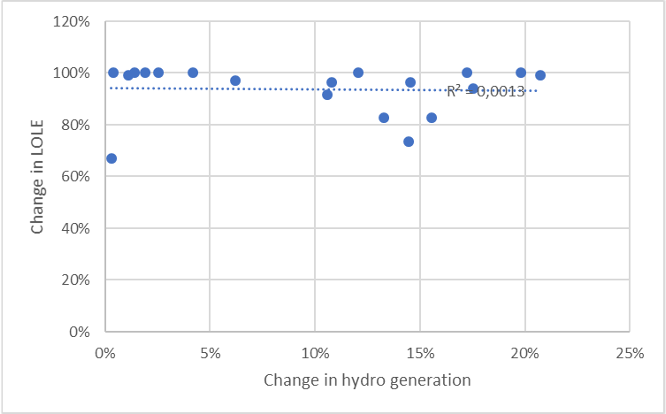}
        \caption{Correlation between percentile change in LOLE and hydro generation (in \% in absolute values) for each market node. }
        \label{SIfig:corr_LOLE-hydro}
\end{figure}

\begin{figure}[h]
        \centering
        \includegraphics[width=.8\textwidth]{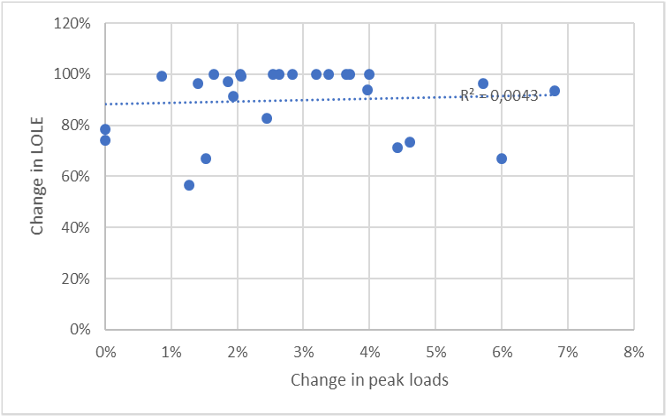}
        \caption{Correlation between percentile change in LOLE and peak loads (in \% in absolute values) for each market node. }
        \label{SIfig:corr_LOLE-load}
\end{figure}

\begin{figure}[h]
        \centering
        \includegraphics[width=.8\textwidth]{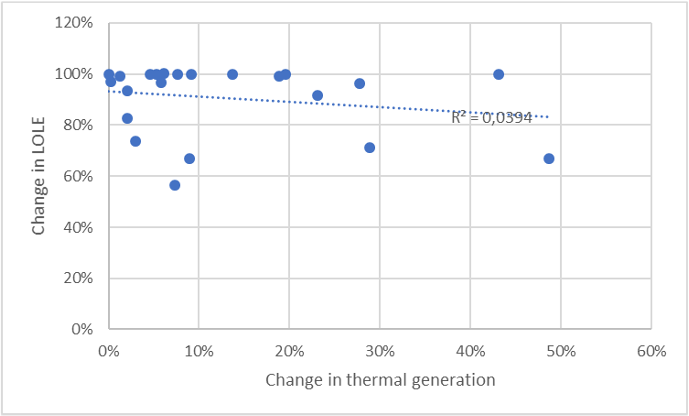}
        \caption{Correlation between percentile change in LOLE and thermal generation (in \% in absolute values) for each market node. }
        \label{SIfig:corr_LOLE-thermal}
\end{figure}

\end{document}